# Forensics Model Cards Generator, V1

**Paola Di Maio**

Independent Analyst, Ronin Institute and Chair, W3C AI Knowledge Representation Community Group

## Abstract

This paper introduces a standardized model card framework specifically designed for digital and web forensics. Building upon established model card methodologies and recent work on abstract models for digital forensic analysis, this paper presents a web based framework that generates model cards specifically designed to represent knowledge in the forensic domain. The framework includes controlled vocabularies for classification, reasoning types, bias identification, and error categorization, along with a web-based generator tool to facilitate adoption. The paper describes the model card structure, presents the controlled vocabularies, and introduces the beta version of the generator tool, inviting community feedback to refine this emerging standard.

**Public Seminar recording**  https://tinyurl.com/publicseminardfmc

# 1. Introduction

## 1.1 Background

Model cards have emerged as a critical framework for documenting machine learning models, promoting transparency, accountability, and responsible AI deployment [1]. Originally proposed by Mitchell et al. (2019), model cards serve as standardized documentation that communicates a model's capabilities, limitations, and appropriate use cases to diverse stakeholders including developers, researchers, policymakers, and end-users.

The representation of Digital forensics processes presents unique challenges due to its complexity especially when coupled with the complexity of AI/ML adoption. Forensic investigations require rigorous documentation to maintain chain of custody, support legal proceedings, and enable reproducibility. The integration of AI/ML systems into forensic workflows introduces additional complexity: models must be explainable, their limitations well-understood, and their outputs defensible in court. Traditional model card frameworks, while valuable, do not adequately capture the specialized requirements of forensic analysis.

## 1.2 Motivation

Recent work by Hargreaves et al. (2024) [2] established an abstract model for digital forensic analysis tools, providing a foundation for systematic error mitigation analysis. Their work identifies key data types and analytical processes specific to digital forensics, creating a structured approach to understanding forensic tool capabilities and limitations.

Building on this foundation, a comprehensive model card framework captures the full context of forensic AI/ML deployments, documents reasoning methodologies specific to investigative work, helps identify sources of bias and error, maps to established forensic workflows and data types, and remains accessible to both technical and non-technical stakeholders. The Digital Forensics Model Card (DF-MC) framework integrates concepts from established practices in traditional model cards while meeting the domain-specific requirements of forensic investigation

## 2. Model Card Structure

The DF-MC framework organizes information into six logical sections, each serving distinct documentation purposes while maintaining coherence across the investigative lifecycle.

### 2.1 Section 1: Identification & Context

This section establishes the model card's identity and basic metadata:

- **MMCID (Model Card Identifier):** Unique identifier following the format DF-MC-YYYY-NNN • **MCV (Version):** Version number for tracking model card evolution
- **DF-MCO (Owner):** Organization or individual responsible for the model

- **DF-MCUse (Usage Context):** Deployment mode (standalone, integrated, or hybrid) • **DF-MC Ln (Layer/Stage):** Position within multi-layered forensic pipelines

This metadata enables precise referencing, version control, and proper attribution—essential requirements for forensic documentation that may be referenced in legal proceedings.

### 2.2 Section 2: Case Context

Forensic investigations are inherently case-specific. This section documents:

- **DF-MC CS (Case Statement):** Investigation scope, objectives, and context • **DF-MC H (Hypothesis):** Investigative hypotheses being tested or explored

By explicitly documenting case context and hypotheses, the model card supports transparent reasoning and enables others to understand the investigative framework within which the AI/ML system operates.

### 2.3 Section 3: Classification & Approach

This section employs controlled vocabularies to categorize the forensic domain and

methodological approach:

- **DF-MC C (Classification):** Forensic domain(s) (e.g., Computer Forensics, Network Forensics, Mobile Device Forensics)

- **DF-MC TR (Type of Reasoning):** Reasoning methodology (e.g., Deductive, Inductive, Abductive, Retroductive)

These classifications enable systematic organization of forensic model cards and facilitate comparison across different implementations and domains.

### 2.4 Section 4: Quality & Limitations

Transparency about limitations is crucial for responsible AI deployment, particularly in forensic contexts where errors can have serious consequences. This section documents:

- **DF-MC B (Bias):** Identified bias types in the system

- **DF-MC CB (Cause of Bias):** Root causes of identified biases

- **DF-MC E (Error):** Observed errors during analysis

- **DF-MC CE (Cause of Error):** Root causes of errors

By systematically documenting bias and error, this section supports critical evaluation of model outputs and helps investigators understand the boundaries of reliable operation.

### 2.5 Section 5: Top Level Elements (DF MC 0)

This section captures high-level conceptual elements of the forensic analysis:

- Algorithm

- Inference methodology

- Confounding factors

- Evaluation approach

- Tools employed

- Evidence handling (MC1)
- File types processed
- Data structures
- Degree of confidence

These elements provide a conceptual map of the analytical framework, enabling stakeholders to understand the overall approach without requiring deep technical knowledge.

### 2.6 Section 6 Data Types & Analytical Processes (DF MC 1)

Building on prior art (Hargreaves et al. 2024), this section documents the specific data processing pipeline:
- Event/Data handling
- Raw data parsing
- Data validation
- Partition identification
- File system processing
- Content identification (carving)
- File type identification
- File-specific processing
- Hashing operations
- Hash matching
- Signature detection
- Timeline construction and analysis
- Geolocation processing and analysis

- Keyword indexing and searching

- Automated result interpretation

- AI-based content flagging

This detailed process documentation enables reproducibility, facilitates error tracing, and supports systematic evaluation of forensic workflows.

## 3. Controlled Vocabularies

Controlled vocabularies are essential for standardized documentation, enabling systematic comparison, aggregation, and analysis across different implementations. The DF-MC framework includes four primary controlled vocabularies.

### 3.1 Forensic Classification Taxonomy

This vocabulary categorizes forensic domains, reflecting the diversity of digital forensic specializations:

- Computer Forensics

- Network Forensics

- Mobile Device Forensics

- Cloud Forensics

- Database Forensics

- Memory Forensics

- Digital Image Forensics

- Digital Video/Audio Forensics

- IoT Forensics

- Multi-domain (covers multiple types)

The taxonomy is based on established forensic subdisciplines [3] and can accommodate emerging domains through the "Other" option.

### 3.2 Reasoning Methodology Taxonomy

Forensic investigation employs diverse reasoning approaches. This vocabulary captures the primary methodologies:

- **Deductive Reasoning:** Proceeding from general principles to specific conclusions
- **Inductive Reasoning:** Drawing general conclusions from specific observations
- **Abductive Reasoning:** Inferring the best explanation for observed evidence
- **Retroductive Reasoning:** Refining and validating hypotheses against gathered data
- **Hybrid/Mixed Reasoning:** Employing multiple reasoning approaches

This taxonomy draws from philosophical logic and forensic science literature [4, 5], recognizing that real world investigations often employ multiple reasoning modes.

### 3.3 Bias Taxonomy

AI bias can significantly impact forensic outcomes. This vocabulary categorizes bias types:

- Data Bias (historical, sampling, selection)

- Algorithmic Bias (model architecture, optimization)

- Human Bias (cognitive, confirmation, implicit)

- Deployment Bias (context mismatch)

- Reporting Bias (documentation gaps)

- Measurement Bias (proxy variables)

- Stereotyping Bias (reinforcing stereotypes)

- Automation Bias (over-reliance on automated results)

- No Identified Bias

- Multiple Bias Types

This taxonomy synthesizes recent work on AI fairness and bias [6, 7], adapted for forensic contexts.

**3.4 Error Causation Taxonomy**

Understanding error sources is critical for improving forensic AI systems. This vocabulary identifies common error causes:
- Training Error (underfitting)

- Validation Error (model selection issues)

- Testing Error (generalization failure)

- Overfitting (high variance)

- Underfitting (high bias)

- Data Quality Issues (noise, outliers, mislabeling)

- Insufficient Training Data

- Class Imbalance

- Feature Engineering Issues

- Hyperparameter Misconfiguration

- Model Complexity Mismatch

- Adversarial Attack (poisoning, evasion)

- Concept Drift

- Tool Calibration Error

- Human Error in Analysis

- Chain of Custody Issues

- Multiple Error Sources

- Unknown/Under Investigation

This taxonomy draws from machine learning literature [8] and forensic practice, recognizing both technical and procedural error sources.

**3.5 Supporting Taxonomies**

Additional controlled vocabularies include:
**DF-MCUse (Usage Context):**

- Standalone

- Integrated

- Hybrid (both standalone and integrated)

**Cause of Bias:**

- Unrepresentative Training Data

- Historical Inequities in Data

- Feature Selection Issues

- Labeling Inconsistencies

- Optimization Objective Mismatch

- Insufficient Diversity in Development Team

- Lack of Domain Expertise

- Temporal Drift (data age/staleness)

- Geographic/Cultural Limitations

- Tool/Method Limitations

- Multiple Causes

- Unknown/Under Investigation

Each controlled vocabulary allows for "Other" entries to accommodate domain-specific variations while maintaining standardization where possible.

## 4. The DF Model Card Generator (Beta)

To facilitate adoption of the DF-MC framework, we have developed a web-based generator tool that guides users through the model card creation process.

### 4.1 Tool Architecture

The generator is implemented as a single-page application using Gradio, providing an accessible web interface that requires no installation. The tool is hosted on Hugging Face Spaces, ensuring open access and reproducibility.

**Key Features:**

1. **Single-Form Interface:** All fields are presented in a logical, scrollable form following the six section structure described above
2. **Controlled Vocabulary Integration:** Dropdown menus and checkbox groups implement the standardized taxonomies
3. **Flexible Documentation:** All fields are optional, accommodating various documentation needs and levels of detail
4. **Dual Output Formats:** Generates both JSON (machine-readable) and Markdown (human readable) outputs
5. **Validation:** Basic format validation for identifier fields while maintaining flexibility

### 4.2 User Workflow

The generator guides users through a systematic documentation process:

1. **Identification:** Enter basic metadata and identifiers

2. **Context:** Document case statement and hypotheses

3. **Classification:** Select forensic domains and reasoning approaches 4. **Quality Assessment:** Document identified biases and errors 5. **Framework Elements:** Check and describe applicable top-level elements 6. **Process Documentation:** Check and describe data processing steps 7. **Generation:** Create model card in JSON and Markdown formats

**4.3 Output Formats**

**JSON Output:** Structured, machine-readable format suitable for:

- Integration with forensic tool pipelines
- Automated analysis and aggregation
- Database storage and retrieval
- API consumption

**Markdown Output:** Human-readable format suitable for:

- Documentation repositories
- Legal proceedings
- Training materials
- Public transparency reports

Both outputs include:
- Complete metadata
- All documented elements
- Generation timestamp
- Generator version

- Schema version

- References to foundational papers

### 4.4 Implementation Details

**Technology Stack:**

- Framework: Gradio 5.x

- Language: Python 3.10+

- Hosting: Hugging Face Spaces

- License: MIT

**Repository:** The generator source code, documentation, and example outputs are available at: https://huggingface.co/spaces/STARBORN/forensics_mc_generator

### 4.5 Current Limitations

This beta release has known limitations:
1. **No Maximum Enforcement:** While guidelines suggest selecting up to 3 items from controlled vocabularies, the tool does not enforce this limit

2. **Limited Validation:** Only MMCID format is validated; other fields accept free-form input
3. **No Versioning Integration:** The tool does not yet integrate with version control systems 4. **Single Language:** Currently available in English only
5. **Basic Search:** No built-in search or filtering of existing model cards

These limitations will be addressed in future releases based on community feedback.

# 5. Beta Testing and Community Feedback

### 5.1 Call for Beta Testers

We invite the digital forensics community to test the DF-MC generator and provide feedback. We are particularly interested in:

**Usability Feedback:**

- Is the workflow intuitive?

- Are the sections logically organized?

- Are field labels and descriptions clear?

- What fields are missing or unnecessary?

**Controlled Vocabulary Feedback:**
- Are the taxonomies comprehensive?

- Do the categories align with forensic practice? • What terms should be added, modified, or removed? • Are there domain-specific vocabularies needed?

**Technical Feedback:**

- Output format requirements

- Integration needs with existing tools

- Performance issues

- Feature requests

**Practical Use Cases:**

- Real-world deployment scenarios

- Example model cards from actual systems • Pain points in current documentation practices • Legal or regulatory requirements

**5.2 How to Contribute**

Feedback can be provided through:

1. **Direct Testing:** Use the generator at https://huggingface.co/spaces/STARBORN/forensics_mc_generator
2. **GitHub Issues:** Report bugs or request features (repository to be announced)
3. **Community Discussion:** Join the W3C Community Group
4. **Email Contact:** Direct correspondence with the author
5. **Example Submissions:** Share generated model cards (with appropriate redaction)

**5.3 Roadmap**

Based on community feedback, we plan to:

**Short-term (3-6 months):**

- Refine controlled vocabularies based on practitioner input
- Implement suggested usability improvements
- Add validation rules and constraints
- Develop example model cards for common forensic tools
- Create documentation and tutorials

**Medium-term (6-12 months):**

- Integration with forensic tool ecosystems
- Version control and model card evolution tracking
- Multi-language support
- Advanced search and filtering capabilities
- API for programmatic model card generation

**Long-term (12+ months):**
- Standards body engagement (ISO, NIST, etc.)
- Formal evaluation of framework effectiveness
- Extension to non-AI forensic tools
- Integration with legal and regulatory frameworks
- Research on model card impact on forensic practice

## 6. Discussion

### 6.1 Benefits of Standardized Documentation

The DF-MC framework offers several benefits to the forensic community:

**Transparency:** Systematic documentation of AI/ML systems builds trust among investigators, legal professionals, and the public.

**Reproducibility:** Detailed process documentation enables independent verification and replication of forensic analyses.

**Quality Assurance:** Structured documentation of bias and error supports continuous improvement and quality control.

**Legal Defensibility:** Comprehensive documentation provides evidence of due diligence and supports expert testimony.

**Knowledge Transfer:** Standardized documentation facilitates training, knowledge sharing, and collaboration.

**Interoperability:** Common frameworks enable comparison across different tools and implementations. **6.2 Challenges and Considerations**
**Documentation Burden:** Creating comprehensive model cards requires time and effort, potentially slowing development cycles.

**Sensitivity Concerns:** Detailed documentation may reveal investigative techniques or system vulnerabilities.

**Dynamic Systems:** Rapidly evolving AI systems may require frequent model card updates.

**Completeness Trade-offs:** Optional fields balance flexibility with comprehensiveness, but may result in incomplete documentation.

**Enforcement:** Without mandates or incentives, adoption may be limited.

### 6.3 Relationship to Existing Standards

### 6.4 Future Directions

**Research Opportunities:**

1. Empirical studies of model card effectiveness in forensic practice
2. Comparative analysis of documented vs. undocumented systems
3. Impact of transparency on legal outcomes
4. Methods for automated model card generation from code
5. Cognitive load and usability studies

**Standards Development:**

1. Engagement with standards bodies (ISO, NIST, SWGDE)
2. Integration with legal frameworks and court requirements
3. Certification programs for compliant systems
4. Industry adoption campaigns

## 7. Conclusion

This paper introduces the Digital Forensics Model Card framework, a standardized approach to documenting AI/ML systems in forensic contexts. By combining established model card principles with domain-specific requirements from digital forensics, the framework provides a structured method for promoting transparency, accountability, and quality in forensic AI deployments.

The accompanying web-based generator tool (beta version) facilitates adoption by providing an accessible interface for creating model cards in both machine-readable and human-readable formats. Controlled vocabularies enable standardized classification and comparison across different implementations.

We invite the digital forensics community to test the generator, provide feedback, and contribute to the evolution of this framework. Through collaborative refinement, we aim to establish a practical standard that serves investigators, developers, legal professionals, and the public interest.

The beta generator is available at:

https://huggingface.co/spaces/STARBORN/forensics_mc_generator **Acknowledgments**

This work builds upon the foundational research of Hargreaves, Nelson, and Casey (2024) on abstract models for digital forensic analysis tools. The author acknowledges the contributions of the W3C AI Knowledge Representation Community Group and the broader digital forensics community in shaping this framework.

**References**

[1] Mitchell, M., Wu, S., Zaldivar, A., Barnes, P., Vasserman, L., Hutchinson, B., Spitzer, E., Raji, I. D., & Gebru, T. (2019). Model Cards for Model Reporting. *Proceedings of the Conference on Fairness, Accountability, and Transparency*, 220-229.

[2] Hargreaves, C., Nelson, A., & Casey, E. (2024). An abstract model for digital forensic analysis tools —A foundation for systematic error mitigation analysis. *Forensic Science International: Digital Investigation*, 48.

[3] Digital Forensics Taxonomy. (2023). *Open University OpenLearn*. Retrieved from https://www.open.edu/openlearn/science-maths-technology/digital-forensics/

[4] Jang, M., Sebire, J., & Stubbs, G. (2025). Detective reasoning in criminal investigation: Integrating abduction, retroduction, deduction, and induction. *Journal of Investigative Psychology and Offender Profiling*.

[5] Pollitt, M. (2008). Applying Traditional Forensic Taxonomy to Digital Forensics. In I. Ray & S. Shenoi (Eds.), *Advances in Digital Forensics IV* (pp. 17-26). Springer.

[6] Mehrabi, N., Morstatter, F., Saxena, N., Lerman, K., & Galstyan, A. (2021). A Survey on Bias and Fairness in Machine Learning. *ACM Computing Surveys*, 54(6), 1-35.

[7] Gallegos, I. O., Rossi, R. A., Barrow, J., Tanjim, M. M., Kim, S., Dernoncourt, F., ... & Ahmed, N. K. (2024). Bias and Fairness in Large Language Models: A Survey. *Computational Linguistics*, 50(3), 1097-1179.

[8] Hastie, T., Tibshirani, R., & Friedman, J. (2009). *The Elements of Statistical Learning: Data Mining, Inference, and Prediction* (2nd ed.). Springer.

## Appendices

See supplementary materials for:

- **Appendix A:** Complete Controlled Vocabularies

- **Appendix B:** Example Model Cards

- **Appendix C:** Generator User Guide

- **Appendix D:** JSON Schema Specification

- **Appendix E:** Integration Guidelines

**Code Repository:**
https://huggingface.co/spaces/STARBORN/forensics_mc_generator **License:** MIT
*attribution/non commercial
**Version:** 1.0-beta (December 2024)